\input harvmac

\def\CN{{\cal N}}

\lref\KL{
V. Kaplunovsky and J. Louis, `On gauge couplings in string theory',
{\it Nucl. Phys.} {\bf B444} (1995) 191, {\tt hep-th/9502077}.}

\Title{\vbox{
\baselineskip12pt\hbox{{\tt hep-th/9608145}}
\hbox{YCTP-P12/96}
}}
{\vbox{
\centerline{Threshold Corrections in $K3 \times T2$ }
\centerline{Heterotic String Compactifications}
}}
\bigskip
\bigskip
\centerline{M{\aa}ns Henningson and Gregory Moore}

\vskip 1cm
\centerline{\it  Department of Physics}
\centerline{\it Yale University}
\centerline{\it P. O. Box 208120}
\centerline{\it New Haven, CT  06520-8120, USA}
\vskip 3mm
\centerline{mans@genesis5.physics.yale.edu}
\centerline{ moore@castalia.physics.yale.edu}

\bigskip
\bigskip

\centerline{\bf Abstract}
We consider compactifications of the heterotic string on $K3 \times T2$ so that
the resulting theory in $d = 4$ space-time dimensions has $N = 2$
supersymmetry. The gravitational and gauge coupling constants of the low-energy
effective theory receive threshold corrections from loops of super-heavy string
states. We calculate these corrections for the case when the
$K3$-surface is a ${\bf Z}_n$ orbifold of a four torus $T4$. The results are
used to determine the one-loop prepotential ${\cal F}_0^{(1)}$ for the vector
multiplets and the gravitational coupling ${\cal F}_1^{(1)}$.

\Date{August 20,1996.}

\newsec{Introduction}
Theories in $d = 4$ space-time dimensions with $N = 2$ supersymmetry have
proven to possess a rich structure while still being simple enough to be
tractable with present methods. In the case of string theory, the gravitational
and gauge coupling constants of the low-energy effective supergravity theory
have received a lot of interest. Their one-loop dependence on the momentum
scale and the moduli is given by
\eqn\stringexpr{
\eqalign{
{1 \over g^2_{\rm grav} (p^2)} & = 24 {\rm Re} \left( -i S + {1 \over 16 \pi^2}
\Delta^{\rm univ} \right) + {b_{\rm grav} \over 16 \pi^2} \log {M^2_{\rm
string} \over p^2} + {1 \over 16 \pi^2} \Delta_{\rm grav}  , \cr
{1 \over g^2_{\rm gauge} (p^2)} & = {\rm Re} \left( -i S + {1 \over 16 \pi^2}
\Delta^{\rm univ} \right) + {b_{\rm gauge} \over 16 \pi^2} \log {M^2_{\rm
string} \over p^2} + {1 \over 16 \pi^2} \Delta_{\rm gauge}  , \cr
}
}
where $S$ is the dilaton   and $\Delta^{\rm univ}$ is a
certain universal contribution related to the Green-Schwarz anomaly
cancellation term. The coefficients $b_{\rm grav}$ and $b_{\rm gauge}$
are related to the one-loop beta-functions and are thus determined
by the spectrum of massless $N = 2$ multiplets \ref\AGN{
I. Antoniadis, E. Gava and K. S. Narain, `Moduli corrections to
gravitational couplings from string loops',
{\it Phys. Lett.} {\bf B283} (1992) 209, {\tt hep-th/9203071} \semi
I. Antoniadis, E. Gava and K. S. Narain, `Moduli corrections to gauge and
gravitational couplings in four-dimensional superstrings',
{\it Nucl. Phys.} {\bf B383} (1992) 93, {\tt hep-th/9204030}.
}\ref\AGNT{
I. Antoniadis, E. Gava, K. Narain and T. Taylor, `Topological amplitudes
in string theory', {\it Nucl. Phys.} {\bf B413} (1994) 162.
}:
\eqn\betaftns{
\eqalign{
b_{\rm grav} & = 46 + 2 (n_H - n_V) \cr
b_{\rm gauge} & = 2 {\rm Tr}_H (Q^2) - 2 {\rm Tr}_V (Q^2) . \cr
}
}
Here $n_H$ and $n_V$ denote the number of
massless hyper multiplets and vector
multiplets respectively, and $Q$ is some generator of the simple factor in
the gauge group under consideration.
The effect of integrating out super-heavy string states is summarized by the
threshold corrections $\Delta_{\rm grav}$ and $\Delta_{\rm gauge}$
\AGN\AGNT\ref\K{
V. Kaplunovsky, 'One-loop threshold effects in string unification',
{\it Nucl. Phys.} {\bf B307} (1988) 145.}\KL.
They are given at one-loop in terms of the fundamental domain integrals
\eqn\Deltadef{
\eqalign{
\Delta_{\rm grav} = \int_{\cal F} {d^2 \tau \over \tau_2} \Biggl[ & {- i
\over
\eta^2 (\tau)} {\rm Tr} \left\{J_0 e^{i \pi J_0}  q^{L_0 - c / 24}
\bar{q}^{\bar{L}_0 - \bar{c} / 24} \left(E_2 (\tau) \! - \! {3 \over \pi
\tau_2}\right) \right\} \! - \! b_{\rm grav} \Biggr] \cr
\Delta_{\rm gauge} = \int_{\cal F} {d^2 \tau \over \tau_2 } \Biggl[ & {- i
\over \eta^2 (\tau)} {\rm Tr} \left\{J_0 e^{i \pi J_0} q^{L_0 - c / 24}
\bar{q}^{\bar{L}_0 - \bar{c} / 24} \left(Q^2 \! - \! {1 \over 8 \pi
\tau_2}\right)
\right\}  - b_{\rm gauge} \Biggr] \cr
}
}
respectively. The traces are over the Ramond sector of the internal conformal
field
theory, and $J_0$ denotes the $U(1)$ generator of the $N = 2$ superconformal
algebra. The functions $\eta$ and $E_2$
are the Dedekind eta function and the Eisenstein series respectively.
Throughout this paper, subscripts $1$ and $2$ on a complex quantity such as
the modular parameter $\tau$ are used to denote its
real and imaginary parts respectively.
Given these threshold corrections, the one-loop prepotential
${\cal F}_0^{(1)}$ for the vector multiplets and the gravitational coupling
${\cal F}_1^{(1)}$  can be determined by comparing the string theory
expressions \stringexpr\ and \Deltadef\ with equations
arising from considerations of holomorphicity and duality transformation
properties in $N = 2$ supersymmetric field theories \KL\ref\dWKLL{
B. de Wit, V. Kaplunovsky, J. Louis and D. L\"ust, `Perturbative couplings
of vector multiplets in $N = 2$ heterotic string vacua',
{\it Nucl. Phys.} {\bf B451} (1995) 53, {\tt hep-th/9504006}.
}
\ref\AFGNT{
I. Antoniadis, S. Ferrara, E. Gava, K. Narain and T. Taylor, `Perturbative
prepotential and monodromies in $N = 2$ heterotic superstring',
{\it Nucl. Phys.} {\bf B447} (1995) 35, {\tt hep-th/9504034}.
}.

We will be considering compactifications of the ten-dimensional heterotic
string to four space-time dimensions with $N = 2$ extended space-time
supersymmetry. This means that the right-moving internal $\bar{c} = 9$
conformal field theory must decompose into a $\bar{c} = 3$ piece and a $\bar{c}
= 6$ piece with $N = 2$ and $N = 4$ world-sheet supersymmetry respectively
\ref\BD{
T. Banks and L. Dixon, `Constraints on string vacua with space-time
supersymmetry',
{\it Nucl. Phys.} {\bf B307} (1988) 93.
}. The
geometrical interpretation is that the six-dimensional internal space is a
direct product of a two-torus $T2$ and a $K3$-surface.
Suppose the gauge bundle on the K3 surface
leaves a rank $s$ subgroup of $E_8 \times E_8$
unbroken. Wilson line
and toroidal moduli then lead to a Narain moduli
space $\CN^{s+2,2}$ for vectormultiplets.
Here
$\CN^{s+2,2}=O(\Gamma^{s+2,2})\backslash
O(s+2,2;R)/K$, where $O(\Gamma^{s+2,2})$ is a
discrete automorphism group of the lattice and
$K$ is the maximal compact subgroup.
Our main object will be to calculate the
threshold corrections $\Delta_{\rm gauge}$
and $\Delta_{\rm grav}$ as functions of these
moduli. Such calculations are feasible essentially
because the threshold corrections only depend on
the elliptic genus of the $\bar{c} = 6$ theory and therefore
can be performed in some convenient limit where this
theory becomes free, such as an orbifold limit.

The standard embedding of the spin connection in the gauge group
generically breaks $E_8\times
E_8 \rightarrow E_7  \times E_8$. The $E_7$ can
be completely Higgsed leaving a rank $12$ group.
The vectormultiplet moduli space is
then $\CN^{s+2,2}$ with $s=8$.
The calculation of $\Delta_{\rm gauge}$ and ${\cal F}_0^{(1)}$ was performed
recently in this case for a ${\bf Z}_2$ orbifold in \ref\HM{
J. Harvey and G. Moore, `Algebras, BPS states and strings',
{\it Nucl. Phys.} {\bf B463} (1996) 315, {\tt hep-th/9510182}.
} by generalizing techniques for performing fundamental domain
integrals introduced in \ref\DKL{
L. Dixon, V. Kaplunovsky and J. Louis,
`Moduli dependence of string loop corrections to gauge coupling constants',
{\it Nucl. Phys.} {\bf B355} (1991) 649.
}.  Ref. \HM\ also
studied in detail the subspace with unbroken $E_8$
corresponding to  $s=0$.   The first gravitational
coupling was evaluated using the same methods
for a special case $s=0$
in \ref\CCLMR{
G. L. Cardoso, G. Curio, D. L\"ust, T. Mohaupt and S.-J. Rey,
`BPS spectra and non-perturbative gravitational couplings in
$N = 2, 4$ supersymmetric string theories',
{\it Nucl. Phys.} {\bf B464} (1996) 18, {\tt hep-th/9512129}.
}.
The methods used in \HM\ apply to a wider
class of backgrounds than the special
cases examined in \HM\CCLMR, and
in this paper we carry out the calculations
for a class of  $Z_n$ orbifolds. The
results of \HM\CCLMR\ left room
for generalization even for the standard
embedding.  In this case one can
begin with the
group $E_7 \times SU(2) \times E_8$ in
the orbifold limit and consider turning
on Wilson lines going to a Coulomb
branch of $E_7 \times E_8$ but leaving
the $SU(2)$
unbroken. We compute the prepotential
on the resulting
vectormultiplet moduli
space $\CN^{17,2}$  by computing
the threshhold correction to the $SU(2)$
coupling. We can then Higgs the $SU(2)$
by turning on K3 moduli.
{}From the calculations below we can
also find the
prepotential and gravitational coupling
for other  topologies of the gauge group.

 It should be noted that
these methods apply in principle to
an even larger class of backgrounds.
\foot{In an intriguing set of papers
\ref\kawai{
T. Kawai, `$N = 2$ heterotic string threshold correction,
$K3$-surface and generalized Kac-Moody superalgebra',
{\it Phys. Lett.} {\bf B372} (1996) 59, {\tt hep-th/9512046}.
}
\ref\neuman{C.D.D Neumann, `The elliptic genus of Calabi-Yau 3- and 4-folds,
product formulae and generalized Kac-Moody algebras',
{\tt hep-th/9607029}. }\
the same methods were applied to obtain interesting
product formulae for automorphic forms, such as the
product formula for the
Siegel automorphic form $\Delta_5$, due to \ref\GN{
V.A. Gritsenko and V. V. Nikulin, `Siegel automorphic form corrections of some
Lorentzian Kac-Moody algebras', {\tt hep-th/9504006}.
}. Unfortunately,
the relevance of the expressions in \kawai\neuman\ to
physical threshhold corrections is unclear. Nevertheless,
the results have been applied in a very interesting
way to black hole physics
\ref\dvv{R. Dijkgraaf, E. Verlinde, and H. Verlinde,
`Counting Dyons in N=4 String Theory',  {\tt hep-th/9607026}.  }. }
For example,
suppose that the lattice $\Gamma^{s+2,2}\in \CN^{s+2,2}$
is $SO(s+2,2)$-related to a lattice of the
form
$\Gamma^{s+2,2} = \amalg [\tilde\Gamma+ \delta_i] $,
where  $\tilde \Gamma \cong \Gamma_0^{s,0} \oplus II^{2,2}$
for some even rank $(s,0)$ lattice $\Gamma^{s,0}_0$
and the $\delta_i$ are orthogonal to $II^{2,2}$.
The modular invariant
integrands of \Deltadef\
take the form
\eqn\form{
\sum_i \Biggl\{ \sum_{p\in \tilde \Gamma + \delta_i}
q^{\half p_L^2} \bar q^{\half p_R^2}
\Biggr\} f_i(q)
}
where $f_i(q)$ form a representation of
the modular group of weight $-s/2$.
Under such conditions the method of
\DKL\ continues to apply with the result that
the threshold correction
$\Delta$ can be decomposed into four
distinct contributions:
\eqn\decomp{
\Delta  = \Delta^{\rm log} + \Delta^{\rm const} + \Delta^{\rm rat} +
\Delta^{\rm transc} .
}
Here $ \Delta^{\rm log}$ and $\Delta^{\rm const}$ are
related to the K\"ahler potential of the moduli space
in a simple way.
$\Delta^{\rm transc} $ is an interesting
sum of polylogarithms of exponentials
of the moduli, weighted by the
Fourier coefficients $c_i(x)$ of $f_i(q)$.
The most subtle term is $\Delta^{\rm rat} $,
 which is a rational function of the moduli.
In order to obtain `nice' automorphic forms
(for example, infinite products with a Weyl
vector)
in the threshhold corrections
it is necessary that the rational terms obey
certain non-trivial identities.
It is far from obvious that this will happen in
general, but we show in section four that
these identities indeed follow from $N = 2$
supersymmetry. Our main results are the formulas
(4.1) - (4.6) for the running coupling
constants and  the expressions (4.13) and (4.14)
for the prepotential and the gravitational coupling
respectively.

The outline of this paper is as follows: In section two, we present the
calculation of the
threshold corrections in the case of ${\bf Z}_n$ orbifolds. In section three,
we specialize the results to the fundamental chamber of the moduli
space. The one-loop prepotential for the vector multiplets and the
gravitational coupling are calculated in section four. The details of the
calculation in section two are discussed in Appendix A. In Appendix B, we
consider the case of the standard embedding of the spin connection in the
gauge group in somewhat more detail.

While this manuscript was being typed, a paper
by T. Kawai appeared
\ref\kawaiii{T. Kawai, `String duality and
modular forms', {\tt hep-th/9607078}. }
which has some overlap with the present paper.

\newsec{The threshold corrections for ${\bf Z}_n$ orbifolds}
We consider $K3$-surfaces which can be obtained as ${\bf Z}_n$ orbifolds of
some four-torus $T4$. Introducing complex coordinates $z^1, z^2$ for $T4$, we
let the group ${\bf Z}_n$ act as
\eqn\groupaction{
\eqalign{
z^1 & \rightarrow e^{2 \pi i a / n} z^1 \cr
z^2 & \rightarrow e^{-2 \pi i a / n} z^2 \cr
}
}
for $a \in {\bf Z} \; {\rm mod} \; n$. Tori which admit such a ${\bf Z}_n$
symmetry exist for $n = 2, 3, 4, 6$. As usual, the Hilbert space of the theory
decomposes into different twist sectors labeled by $a = 0, \ldots, n - 1$. We
will also have to project onto ${\bf Z}_n$ invariant states in the partition
function, which is equivalent to inserting a group element indexed by $b$ in
the `time' direction, summing over $b = 0, \ldots, n - 1$ and dividing by $n$.

In the bosonized formulation of the heterotic string, the gauge degrees of
freedom are described by sixteen left-moving bosons. Together with the two
left-moving and two right-moving bosons from the two-torus $T2$, these fields
take their values on a torus given by an even self-dual lattice $\Gamma^{18,
2}$ of the indicated signature. To cancel the space-time anomaly, we let ${\bf
Z}_n$ act by a shift ${a \over n} \gamma$ on the momentum, where $\gamma \in
\Gamma^{18, 2}$. The contribution to the integrands in \Deltadef\ in the $(a,
b)$ sector from these degrees of freedom is then
\eqn\Zlattice{
Z^{\rm torus}_{a, b} (\tau, \bar{\tau}) = 2 e^{- 2 \pi i {a b \over n^2}
\gamma^2}
\eta^{-18} (\tau) \sum_{p \in
\Gamma^{18, 2} + {a \over n} \gamma} e^{2 \pi i {b \over n} p \cdot \gamma}
q^{{1 \over 2} p_L^2} \bar{q}^{{1 \over 2} p_R^2}
}
for $\Delta_{\rm grav}$ and
\eqn\Zhat{
\eqalign{
\hat{Z}^{\rm torus}_{a, b} (\tau, \bar{\tau}) = & e^{- 2 \pi i {a b \over n^2}
\gamma^2}
\eta^{-18} (\tau) \cr
& \times \sum_{p \in
\Gamma^{18, 2} + {a \over n} \gamma} e^{2 \pi i {b \over n} p \cdot \gamma}
q^{{1 \over 2} p_L^2}
\bar{q}^{{1 \over 2} p_R^2} \left( {(p \cdot Q)^2 \over Q \cdot Q} - {1 \over 4
\pi \tau_2} \right)
}
}
for $\Delta_{\rm gauge}$. Notice that there are no right-moving oscillator
contributions
because of a cancellation between bosons and fermions. It is easy to check that
\eqn\Zlatticeperiod{
\eqalign{
Z^{\rm torus}_{a, b + n} (\tau, \bar{\tau}) & = Z^{\rm torus}_{a, b} (\tau,
\bar{\tau}) \cr
\hat{Z}^{\rm torus}_{a, b + n} (\tau, \bar{\tau}) & =
\hat{Z}^{\rm torus}_{a, b} (\tau, \bar{\tau}) \cr
}
}
and that
\eqn\Zlatticemodular{
\eqalign{
Z^{\rm torus}_{a, b} (\tau + 1, \bar{\tau} + 1) & =
\exp 2 \pi i \left({1 \over 2} {a^2 \over n^2} \gamma^2 - {3 \over 4} \right)
Z^{\rm torus}_{a, a + b} (\tau, \bar{\tau}) \cr
Z^{\rm torus}_{a, b} (- 1 / \tau, - 1 / \bar{\tau}) & =
\exp 2 \pi i \left({a b \over n^2} \gamma^2 \right)
 (i \bar{\tau})^{1/2} Z^{\rm torus}_{b, -a} (\tau, \bar{\tau}) \cr
\hat{Z}^{\rm torus}_{a, b} (\tau + 1, \bar{\tau} + 1) & =
\exp 2 \pi i \left({1 \over 2} {a^2 \over n^2} \gamma^2 - {3 \over 4} \right)
\hat{Z}^{\rm torus}_{a, a + b} (\tau, \bar{\tau}) \cr
\hat{Z}^{\rm torus}_{a, b} (- 1 / \tau, - 1 / \bar{\tau}) & =
\exp 2 \pi i \left({a b \over n^2} \gamma^2 \right) (-i \tau) (i
\bar{\tau})^{1/2}
\hat{Z}^{\rm torus}_{b, -a} (\tau, \bar{\tau}) . \cr
}
}
The modular transformation laws for $\hat{Z}^{\rm torus}_{a, b} (\tau,
\bar{\tau})$
require that the charge
$Q$ be purely left-moving. Physically, this means that the unbroken gauge group
contains at least an $SU(2)$ factor.

Next, we consider the contributions to the integrands in \Deltadef\ from the
degrees
of freedom associated with the $K3$ surface. We begin with untwisted sector,
where
$a = 0$. The contributions from the non-zero
modes of the right-moving bosons and fermions cancel against each other, so we
only get a factor $16 \sin^2 \pi {b \over n}$
from the fermionic zero-modes. This is the well known holomorphicity of the
elliptic genus.
The contribution from the left-movers is
\eqn\leftkthree{
q^{-{1 \over 6}} \prod_{j = 1}^\infty \left(1 - e^{2 \pi i {b \over n}} q^j
\right)^{-2}
\left(1 - e^{-2 \pi i {b \over n}} q^j \right)^{-2} .
}
The contributions in the twisted sectors, where $a \neq 0$, are now determined
by
the requirement that the integrands
\eqn\gravintegrand{
\eqalign{
{-i \tau_2 \over \eta^2 (\tau)} & {\rm Tr} \left\{J_0 e^{i \pi J_0} q^{L_0 - c
/ 24} \bar{q}^{\bar{L}_0 - \bar{c} / 24} \left(E_2 (q) - {3 \over \pi
\tau_2}\right) \right\} \cr
& = {\tau_2 \over \eta^2 (\tau)} \sum_{a, b} {1 \over n} Z^{K3}_{a, b} (\tau)
Z^{\rm torus}_{a, b} (\tau, \bar{\tau}) \left(E_2 (q) - {3 \over \pi
\tau_2}\right) \cr
}
}
and
\eqn\gaugeintegrand{
\eqalign{
{- i \tau_2 \over \eta^2 (\tau)} & {\rm Tr} \left\{J_0 e^{i \pi J_0} q^{L_0 - c
/ 24} \bar{q}^{\bar{L}_0 - \bar{c} / 24} \left(Q^2 - {1 \over 8 \pi
\tau_2}\right) \right\} \cr
& = {\tau_2 \over \eta^2 (\tau)} \sum_{a, b} {1 \over n} Z^{K3}_{a, b} (\tau)
\hat{Z}^{\rm torus}_{a, b} (\tau, \bar{\tau}) \cr
}
}
be modular invariants. In general, we can write
\eqn\kthree{
Z^{\rm K3}_{a, b} (\tau) = k_{a, b} \; q^{- \left({a \over n} \right)^2} \eta^2
(\tau)
\Theta_1^{-2} \left( \tau {a \over n} + {b \over n} \bigl| \tau \right) ,
}
where $\eta (\tau)$ and $\Theta_1 (\nu | \tau)$ are the Dedekind eta function
and the Jacobi theta function respectively and $k_{a, b}$ are some constants.
{}From the above, it follows that in the untwisted sector
\eqn\untwisted{
k_{0, b} = 64 \sin^4 \pi {b \over n} .
}
The $k_{a, b}$ for $a \neq 0$ are now determined by modular invariance through
the
relations
\eqn\modularconditions{
\eqalign{
{k_{a, b} \over k_{a, a + b}} & = e^{i \pi {a^2 \over n^2} \left( 2 - \gamma^2
\right)} \cr
{k_{a, b} \over k_{b, -a}} & = e^{-2 \pi i {a b \over n^2} \left( 2 - \gamma^2
\right)} . \cr
}
}
These relations are only consistent provided that the level matching condition
\eqn\levelmatch{
\gamma^2 - 2 \in 2 n {\bf Z}
}
is fulfilled.

The $\Gamma^{18, 2}$ lattice is obtained by an $SO(18, 2)$ rotation of some
standard lattice, which we take to be of the form $II^{16, 0} \oplus II^{1,
1} \oplus II^{1,1}$. Here $II^{16, 0}$ is a sixteen-dimensional, even self
dual Euclidean lattice, i.e. either the $E_8 \times E_8$ root lattice or the
${\rm Spin} (32) / {\bf Z}_2$ weight lattice. The shift $\gamma$ and the charge
$Q$ transform as vectors under $SO(18, 2)$. In the standard lattice, we take
$\gamma$ to be given by a lattice vector $\bar{\gamma}$ in the $II^{16, 0}$
factor.
\foot{A bar over a vector, as in $\bar{\gamma}$
indicates the component in the $II^{16, 0}$ factor.}
The unbroken gauge group is
then generated by the root vectors $\bar{r}$ of $II^{16, 0}$ which fulfil ${1
\over n} \bar{r} \cdot \bar{\gamma} \in {\bf Z}$. In the standard lattice, $Q$
equals one of these root vectors, which we denote by $\bar{Q}$.

The subgroup $SO(18) \times SO(2)$ gives rise to equivalent lattices, so the
moduli space is the quotient by the discrete $T$-duality group of
\eqn\modulispace{
  SO(18, 2) / (SO(18) \times SO(2)),
}
which may be parametrized by complex moduli $y = (\bar{y}, y^+, y^-)$ subject
to the restrictions $y_2^- > 0$ and $(y_2 , y_2) < 0$ \dWKLL\ref\CDFV{
A. Ceresole, R. D'Auria, S. Ferrara and A. van Proeyen,
`On electromagnetic duality in locally supersymmetric $N = 2$ Yang-Mills
theory',
{\tt hep-th/9412200}.
}. Here the inner product
is given by
\eqn\product{
(y, y^\prime) = \bar{y} \cdot \bar{y}^\prime - y^+ y^{-\prime} - y^-
y^{+\prime},
}
where $\bar{y} \cdot \bar{y}^\prime$ denotes the standard Euclidean product
between two sixteen dimensional vectors. The moduli space is a K\"ahler space
with K\"ahler potential
\eqn\Kahler{
K = - \log \bigl( - (y_2, y_2) \bigr) .
}
In these coordinates, the left- and right-moving components of $p \in
\Gamma^{18, 2} + {a \over n} \gamma$ are given by
\eqn\momenta{
\eqalign{
{1 \over 2} p_L^2 - {1 \over 2} p_R^2 & = {1 \over 2} \bar{R} \cdot \bar{R} -
m^+ n^- + m^0 n^0 \cr
{1 \over 2} p_R^2 & = {1 \over -2 (y_2, y_2)} \left| \bar{R} \cdot \bar{y} +
m^+ y^- + n^- y^+ + m^0 + {1 \over 2} n^0 (y, y) \right|^2, \cr
}
}
where $\bar{R} = \bar{r} + {a \over n} \bar{\gamma}$ and $\bar{r} \in II^{16,
0}$ and
$m^+, n^-, m^0, n^0 \in {\bf Z}$ label a vector in the standard lattice. We see
that the requirement that the charge $Q$ be purely left-moving is equivalent to
demanding that $\bar{y} \cdot \bar{Q} = 0$. This means that the $Q$ gauge boson
is massless so that  the unbroken gauge group is at least $SU(2)$.

The constants $b_{\rm grav}$ and $b_{\rm gauge}$ are determined by the
requirement that $\Delta_{\rm grav}$ and $\Delta_{\rm gauge}$ be finite at a
generic point in the moduli space and are thus given by the coefficients of
$\tau_2 q^0 \bar{q}^0$ in the integrands. Introducing the Fourier coefficients
$c_{a, b} (h)$ and $\tilde{c}_{a, b} (h)$ through
\eqn\Fourier{
\eqalign{
e^{-2 \pi i {a b \over n^2} \gamma^2} \eta^{-20} (\tau) Z^{K3}_{a, b} (\tau) &
= \sum_{h \geq -1} c_{a, b} (h) q^h \cr
e^{-2 \pi i {a b \over n^2} \gamma^2} \eta^{-20} (\tau) E_2 (\tau) Z^{K3}_{a,
b} (\tau) &
= \sum_{h \geq -1} \tilde{c}_{a, b} (h) q^h . \cr
}
}
we get
\eqn\bvalues{
\eqalign{
b_{\rm grav} & = \sum_{\bar{r}} \sum_{a, b}{}^{\! {}^0} {1 \over n}
e^{2 \pi i {b \over n} \bar{R} \cdot \bar{\gamma}} \tilde{c}_{a, b} \left( -{1
\over 2} \bar{R}
\cdot \bar{R} \right) 2 \cr
b_{\rm gauge} & = \sum_{\bar{r}} \sum_{a, b}{}^{\! {}^0} {1 \over n}
e^{2 \pi i {b \over n} \bar{R} \cdot \bar{\gamma}} c_{a, b} \left( -{1 \over 2}
\bar{R} \cdot
\bar{R} \right) {(\bar{R} \cdot \bar{Q})^2 \over \bar{Q} \cdot \bar{Q}} . \cr
}
}
The superscript $0$ on the summation signs indicate that, for a given
$\bar{r}$,
the sum is only over $a$ such that $\bar{R} \cdot \bar{y} = 0$ for
generic values of the moduli $\bar{y}$ in the moduli space. In terms of the
massless
spectrum, we get \betaftns. The sums in \Fourier\
should be regarded as sums over the real numbers,
with $c_{a,b}(x)$ only taking nonzero values on the
discrete spectrum of the model.

To write the final answer for $\Delta_{\rm grav}$ and $\Delta_{\rm gauge}$, we
make the following definitions:
First we introduce a notion of positivity of the vectors $\bar{R} = \bar{r}
+ {a \over n} \bar{\gamma}$. (In the case of $\bar{\gamma}$ being a root
vector, corresponding
to the standard embedding of the spin connection in the gauge group, this can
be done by first defining a positivity condition on lattice vectors $\bar{r}
\in II^{16, 0}$ such that $\bar{\gamma}$ is the smallest positive root. We then
define $\bar{R}$ to be positive if  $\bar{r} > 0$ or $\bar{r} = 0$ and $a >
0$.) We now say that the triplet $R = (\bar{R}, -k, -l)$ is positive if
\eqn\Rpos{
\eqalign{
& k > 0 \;\; {\rm or} \cr
& k = 0, \;\; l > 0 \;\; {\rm or} \cr
& k = l = 0, \;\; \bar{R} > 0 . \cr
}
}
It is convenient to introduce the functions
\eqn\ddef{
\eqalign{
d ( R ) & \equiv \sum_b {1 \over n} e^{2 \pi i {b \over n} \bar{R} \cdot
\bar{\gamma}}
c_{a, b} \left( - {1 \over 2} (R, R) \right) \cr
\tilde{d} ( R ) &  \equiv \sum_b {1 \over n} e^{2 \pi i {b \over n} \bar{R}
\cdot \bar{\gamma}}
\tilde{c}_{a, b} \left( - {1 \over 2} (R, R) \right) . \cr
}
}
We will also use these functions with argument $\bar{R}$ instead of $R$,
meaning
that $k = l = 0$.
We define the product $(R ; y)$ as
\eqn\semiproduct{
(R ; y) = \cases{
\bar{R} \cdot \bar{y}_1 + l y^-_1 + k y^+_1 + i \left| \bar{R} \cdot \bar{y}_2
+ l y^-_2 + k y^+_2 \right| & for $k > 0$ \cr
\bar{R} \cdot \bar{y} + l y^- - \left[ {\bar{R} \cdot \bar{y}_2 \over y_2^-}
\right] y^- & for $k = 0$, $\bar{R} \geq 0$ \cr
\bar{R} \cdot \bar{y} + l y^- + \left[ -{\bar{R} \cdot \bar{y}_2 \over y_2^-}
\right] y^- & for $k = 0$, $\bar{R} < 0$ , \cr
}
}
where $[x]$ is the greatest integer less than or equal to $x$.  Finally,
we introduce the functions
\eqn\polylog{
\eqalign{
{\rm li}_m (x) & = {\rm Li}_m (e^{2 \pi i x}) = \sum_{p = 1}^\infty
{\left(e^{2 \pi i x} \right)^p \over p^m} \cr
{\cal P} (x) & = x_2 {\rm li}_2 (x) + {1 \over 2 \pi} {\rm li}_3 (x) . \cr
}
}

The calculation of $\Delta_{\rm grav}$ and $\Delta_{\rm gauge}$
 is performed in Appendix A. The result is
\eqn\Deltadecomp{
\eqalign{
\Delta_{\rm grav} & = \Delta_{\rm grav}^{\rm log} + \Delta_{\rm grav}^{\rm
const} + \Delta_{\rm grav}^{\rm rat} + \Delta_{\rm grav}^{\rm transc} \cr
\Delta_{\rm gauge} & =  \Delta_{\rm gauge}^{\rm log} + \Delta_{\rm gauge}^{\rm
const} + \Delta_{\rm gauge}^{\rm rat} + \Delta_{\rm gauge}^{\rm transc}, \cr
}
}
where
\eqn\grterms{
\eqalign{
\Delta_{\rm grav}^{\rm log} = & \; b_{\rm grav} \left( - \log \bigl(-(y_2, y_2)
\bigr) - {\cal K} \right) \cr
\Delta_{\rm grav}^{\rm const} = & {96 \zeta (3) \chi \over \pi^2 (y_2, y_2)}
\cr
\Delta_{\rm grav}^{\rm rat} = & \sum_{\bar{r}, a} \Biggl( \tilde{d} (\bar{R})
 {4 y^-_2 \over \pi} {\rm Re} \; {\rm li}_2 \left({\bar{R}
\cdot \bar{y}_2 \over y_2^-} \right) \cr
& \;\;\;\;\;\;\; + d (\bar{R}) \left( -{\pi (y_2, y_2) \over 5 y^-_2} + {24
(y_2^-)^3
\over \pi^3 (y_2, y_2)}
{\rm Re} \; {\rm li}_4 \left({\bar{R} \cdot \bar{y}_2 \over y_2^-} \right)
\right) \Biggr) \cr
\Delta_{\rm grav}^{\rm transc} = & \sum_{R > 0}{}^{\! {}^\prime} \Biggl(
\tilde{d} (R)
 8 {\rm Re \;} {\rm li}_1 \bigl( (R ; y) \bigr) + d (R) {48 \over \pi (y_2,
y_2)}
{\rm Re \;} {\cal P} \bigl( (R ; y) \bigr) \Biggr) \cr
}
}
and
\eqn\Qterms{
\eqalign{
\Delta_{\rm gauge}^{\rm log} = & \; b_{\rm gauge} \left( - \log \bigl(-(y_2,
y_2)
\bigr) - {\cal K} \right) \cr
\Delta_{\rm gauge}^{\rm const} = & {4 \zeta (3) \chi \over \pi^2 (y_2, y_2)}
\cr
\Delta_{\rm gauge}^{\rm rat} = & \sum_{\bar{r}, a} d (\bar{R}) \Biggl(
 {(\bar{R} \cdot \bar{Q} )^2 \over \bar{Q} \cdot \bar{Q}}
\left( {- \pi (y_2, y_2) \over 6 y^-_2} + {2 y^-_2 \over
\pi}
{\rm Re} \; {\rm li}_2 \left({\bar{R} \cdot \bar{y}_2 \over y_2^-} \right)
\right) \cr
&  \;\;\;\;\;\;\;\;\;\;\;\;\;\;\;\;\; + {\pi (y_2, y_2) \over 120 y^-_2}
+ {(y_2^-)^3 \over \pi^3 (y_2, y_2)} {\rm Re} \; {\rm li}_4
\left({\bar{R} \cdot \bar{y}_2 \over y_2^-} \right) \Biggr) \cr
\Delta_{\rm gauge}^{\rm transc} = & \sum_{R > 0}{}^{\! {}^\prime} d (R) \Biggl(
 {(\bar{R} \cdot \bar{Q})^2 \over \bar{Q} \cdot
\bar{Q}} 4 {\rm Re} \; {\rm li}_1 \bigl( (R ; y) \bigr)
+ {2 \over \pi (y_2, y_2)} {\rm Re} \; {\cal P} \bigl( (R ; y) \bigr) \Biggr).
\cr
}
}
Here ${\cal K} = 1 - \gamma_E +  \log {4 \pi \over 3 \sqrt{3}}$ and
\eqn\chidef{
\chi = {1 \over 4} \sum_{\bar{r}, a}{}^{\! {}^0} d (\bar{R}) .
}
The prime on the sums over $R > 0$ in $\Delta^{\rm transc}_{\rm grav}$ and
$\Delta^{\rm transc}_{\rm gauge}$ indicates that terms with $k = l = 0$ and
$\bar{R} \cdot \bar{y} = 0$ for generic values of the moduli are omitted.
These are exactly the terms which are included in the sum with the superscript
zero in $\chi$.

\newsec{The fundamental chamber}
The above formulas can be simplified in the (generalized) fundamental
chamber of the moduli
space defined by the conditions
\eqn\fundamental{
\eqalign{
& 0 < {\bar{R} \cdot \bar{y}_2 \over y^-_2} < 1  \;\;\;\;  {\rm for} \;\;
\bar{R} > 0, \;\; \bar{R} \cdot \bar{R} \leq 2; \cr
& 0 < y^-_2 < y^+_2 . \cr
}
}
This means that $(R ; y) = (R , y)$ for all $R$ such that $-{1 \over 2} (R, R)
\geq - 1$. Furthermore, using the identities
\eqn\identities{
\eqalign{
{\rm Re} \; {\rm li_2} (x) & = \pi^2 \left( {1 \over 6} - | x | + x^2 \right)
\cr
{\rm Re} \; {\rm li_4} (x) & = \pi^4 \left( {1 \over 90} - {1 \over 3} x^2 + {2
\over 3} | x |^3
- {1 \over 3} x^4 \right) , \cr
}
}
which are valid for $-1 \leq x \leq 1$,
the rational terms can now be written as
\eqn\rational{
\eqalign{
\Delta_{\rm grav}^{\rm rat} & = {24 \pi \over (y_2, y_2)} \left( (d_{\rm
grav})_{ABC} y_2^A y_2^B y_2^C + {1 \over y_2^-} (\hat{d}_{\rm grav})_{ABCD}
y_2^A y_2^B y_2^C y_2^D \right) \cr
\Delta_{\rm gauge}^{\rm rat} & = {\pi \over (y_2, y_2)} \left( (d_{\rm
gauge})_{ABC} y_2^A y_2^B y_2^C + {1 \over y_2^-} (\hat{d}_{\rm gauge})_{ABCD}
y_2^A y_2^B y_2^C y_2^D \right) , \cr
}
}
where
\eqn\dgr{
\eqalign{
(d_{\rm grav})_{ABC} & y_2^A y_2^B y_2^C \cr
= \sum_{\bar{r}, a} \Biggl( & \tilde{d} (\bar{R})
\Bigl( -{1 \over 6} | \bar{R} \cdot \bar{y}_2 |  \bar{y}_2 \cdot
\bar{y}_2  + {1 \over 3} | \bar{R} \cdot \bar{y}_2 | y^+_2 y^-_2 \cr
& \;\;\;\;\;\;\;\;\;\; - {1 \over 3} (\bar{R} \cdot \bar{y}_2)^2 y^+_2 + {1
\over 36} \bar{y}_2
\cdot \bar{y}_2 y_2^- - {1 \over 18} y^+_2 y^-_2 y^-_2 \Bigr) \cr
& + d (\bar{R}) \Bigl( {2 \over 3} | \bar{R} \cdot \bar{y}_2 |^3 + {1 \over 30}
\bar{y}_2 \cdot \bar{y}_2 y^+_2 - {1 \over 3} (\bar{R} \cdot \bar{y}_2)^2 y^-_2
\cr
& \;\;\;\;\;\;\;\;\;\;\;\;\;\; - {1 \over 30} y^+_2 y^+_2 y^-_2
+ {1 \over 90} y^-_2 y^-_2 y^-_2 \Bigr) \Biggr) , \cr
}
}
\eqn\dQ{
\eqalign{
(d_{\rm gauge})_{ABC} y_2^A & y_2^B y_2^C \cr
= \sum_{\bar{r}, a} d (\bar{R}) \Biggl( & {(\bar{R} \cdot \bar{Q})^2 \over
\bar{Q} \cdot
\bar{Q}} \Bigl( -2 | \bar{R} \cdot \bar{y}_2 |  \bar{y}_2 \cdot \bar{y}_2  + 4
| \bar{R} \cdot \bar{y}_2 | y^+_2 y^-_2  + {2 \over 3} \bar{y}_2 \cdot
\bar{y}_2 y^+_2 \cr
& \;\;\;\;\;\;\;\;\;\;\;\;\;  - 4 (\bar{R} \cdot \bar{y}_2)^2 y^+_2
+ {1 \over 3} \bar{y}_2 \cdot \bar{y}_2
y_2^- - {2 \over 3} y^+_2 y^-_2 y^-_2 - {2 \over 3} y^+_2 y^+_2 y^-_2 \Bigr)
\cr
& +  {2 \over 3} | \bar{R} \cdot \bar{y}_2 |^3 - {1 \over 30}
\bar{y}_2 \cdot \bar{y}_2 y^+_2 - {1 \over 3} (\bar{R} \cdot \bar{y}_2)^2 y^-_2
+ {1 \over 30} y^+_2 y^+_2 y^-_2 \cr
& + {1 \over 90} y^-_2 y^-_2 y^-_2 \Biggr) , \cr
}
}
\eqn\dhatgr{
\eqalign{
(\hat{d}_{\rm grav} & )_{ABCD} y_2^A y_2^B y_2^C y_2^D \cr
& = \sum_{\bar{r}, a}  \Biggl( \tilde{d} (\bar{R}) {1 \over 6} (\bar{R} \cdot
\bar{y}_2)^2
\bar{y}_2 \cdot \bar{y}_2
+ d (\bar{R}) \left( -{1 \over 120} (\bar{y}_2 \cdot \bar{y}_2)^2 - {1 \over 3}
(\bar{R} \cdot \bar{y}_2)^4 \right) \Biggr) \cr
}
}
and
\eqn\dhatQ{
\eqalign{
(\hat{d}_{\rm gauge})_{ABCD} & y_2^A y_2^B y_2^C y_2^D \cr
= \sum_{\bar{r}, a} d (\bar{R}) \Biggl(
& {(\bar{R} \cdot \bar{Q})^2 \over \bar{Q} \cdot \bar{Q}}
\left( 2 (\bar{R} \cdot \bar{y}_2)^2 \bar{y}_2 \cdot \bar{y}_2 - {1
\over 6} (\bar{y}_2 \cdot \bar{y}_2)^2 \right) \cr
& + {1 \over 120} (\bar{y}_2 \cdot \bar{y}_2)^2 - {1 \over 3}
(\bar{R} \cdot \bar{y}_2)^4 \Biggr) . \cr
}
}
In fact, as we will see in the next section, these tensors obey
the following identites on the subspace where $\bar{y} \cdot \bar{Q} = 0$:
\eqn\condone{
\eqalign{
(\hat{d}_{\rm grav})_{ABCD} y_2^A y_2^B y_2^C y_2^D & = 0 \cr
 (\hat{d}_{\rm gauge})_{ABCD} y_2^A y_2^B y_2^C y_2^D & = 0  \cr
}
}
and
\eqn\condtwo{
\left( d_{\rm grav} - d_{\rm gauge} \right)_{ABC} y_2^A y_2^B y_2^C = 0 \;\;
{\rm mod} \;\; (y_2, y_2) \tilde{d}_A y_2^A
}
for some real $\tilde{d}_A$.
The last equation obviously implies that the tensor $(d_{\rm gauge} )_{ABC}
y_2^A y_2^B y_2^C$ is independent of the factor of the gauge group under
consideration up to
terms of the form $(y_2, y_2) \tilde{d}_A y_2^A$.
As we show in the next section, it  follows from $N = 2$
supersymmetry that these conditions are
obeyed in general, although this is far from obvious in our calculation.
 In Appendix B, we check that they are
indeed fulfilled in the case of the standard embedding
of the spin connection in the gauge group.

We have thus found that, in the fundamental chamber and on the subspace where
$\bar{y} \cdot \bar{Q} = 0$, the threshold corrections are
\eqn\Deltafund{
\eqalign{
\Delta_{\rm grav} = \; & b_{\rm grav} \left( - \log \left(-(y_2, y_2) \right) -
{\cal K} \right) + {96 \zeta (3) \chi \over \pi^2 (y_2, y_2)} \cr
& + {24 \pi \over (y_2, y_2)} (d_{\rm grav})_{ABC} y^A_2 y^B_2 y^C_2 \cr
& + \sum_{R > 0}{}^{\! {}^\prime} \Biggl( \tilde{d} (R) 8 {\rm Re} \;
{\rm li}_1 \bigl( (R, y) \bigr)
+ d (R) {48 \over \pi (y_2, y_2)} {\rm Re} \; {\cal P} \bigl( (R, y) \bigr)
\Biggr) \cr
\Delta_{\rm gauge} = \; & b_{\rm gauge} \left( - \log \left(-(y_2, y_2) \right)
- {\cal K} \right) + {4 \zeta (3) \chi \over \pi^2 (y_2, y_2)} \cr
& + {\pi \over (y_2, y_2)} (d_{\rm gauge})_{ABC} y^A_2 y^B_2 y^C_2 \cr
& + \sum_{R > 0}{}^{\! {}^\prime} d (R) \Biggl( 4 {(\bar{R} \cdot
\bar{Q})^2 \over \bar{Q} \cdot \bar{Q}} {\rm Re} \; {\rm li}_1 \bigl( (R, y)
\bigr) + {2 \over \pi (y_2, y_2)} {\rm Re} \; {\cal P} \bigl( (R, y) \bigr)
\Biggr) .  \cr
}
}

\newsec{The prepotential and the gravitational coupling}
In this section, we will compute the prepotential for the vector multiplets
${\cal F}_0^{(1)}$ and the gravitational coupling ${\cal F}_1^{(1)}$ and show
that the effective (non-Wilsonian)
coefficients of the operators ${\rm Tr} F^2$ and $R
\wedge R^*$ may be written in the form
\eqn\running{
\eqalign{
{1 \over g^2_{\rm grav} (p^2)} & = 24 {\rm Im} \tilde{S}
+ {b_{\rm grav} \over 16 \pi^2} \log {M_{\rm str}^2 \over p^2}
 - {3 \over 4 \pi^2} \log || \Psi_{\rm grav} ||^2 \cr
{1 \over g^2_{\rm gauge} (p^2)} & = {\rm Im} \tilde{S}
+ {b_{\rm gauge} \over 16 \pi^2} \log {M_{\rm str}^2 \over p^2}
 - {1 \over  (s + 4) 4 \pi^2} \log || \Psi_{\rm gauge} ||^2 , \cr
}
}
where  $\tilde{S}$ is the {\it pseudo-invariant} dilaton defined below
\footnote{${}^1$}{
$\tilde{S}$ differs from the dilaton $S$ by the addition of a holomorphic
function
such that $\tilde{S}$ transforms with a real shift under duality
transformations
\dWKLL\HM. Note that this is slightly different from the {\it invariant}
dilaton of
\dWKLL.
}
and the holomorphic functions $\Psi_{\rm grav}$ and $\Psi_{\rm gauge}$
are automorphic forms of the $T$-duality group. In particular, the invariant
norm
squares are
\eqn\norms{
\eqalign{
|| \Psi_{\rm grav} ||^2 & = \bigl( - (y_2, y_2) \bigr)^{w_{\rm grav}}
| \Psi_{\rm grav} |^2 \cr
|| \Psi_{\rm gauge} ||^2 & =  \bigl( - (y_2, y_2) \bigr)^{w_{\rm gauge}}
| \Psi_{\rm gauge} |^2 \cr
}
}
with the weights
\eqn\weights{
\eqalign{
w_{\rm grav} & = {1 \over 12} b_{\rm grav} \cr
w_{\rm gauge} & = {s + 4 \over 4} b_{\rm gauge} . \cr
}
}
Furthermore, these holomorphic functions may be written in the form
\eqn\productform{
\eqalign{
\Psi_{\rm grav} & = e^{2 \pi i (\rho_{\rm grav})_A y^A}
\prod_{R > 0}{}^{\! {}^\prime} \left( 1 - e^{2 \pi i (R, y)} \right)^{l_{\rm
grav} (R)} \cr
\Psi_{\rm gauge} & = e^{2 \pi i (\rho_{\rm gauge})_A y^A}
\prod_{R > 0}{}^{\! {}^\prime} \left( 1 - e^{2 \pi i (R, y)} \right)^{l_{\rm
gauge} (R)} , \cr
}
}
where the Weyl vectors and exponents are given by
\eqn\Weylvectors{
\eqalign{
(\rho_{\rm grav})_A & = {576 \over s + 4} (d_{\rm grav})^B{}_{BA} \cr
(\rho_{\rm gauge})_A & = 3 (d_{\rm gauge})^B{}_{BA}\cr
}
}
and
\eqn\exponents{
\eqalign{
l_{\rm grav} (R) & = {96 \over s + 4} d (R) \; (R, R) - 8 \tilde{d} (R) \cr
l_{\rm gauge} (R) & = d (R) \left( {1 \over 2} (R, R)
- {s + 4 \over 2} {(\bar{R} \cdot \bar{Q})^2 \over \bar{Q} \cdot \bar{Q}}
\right) \cr
}
}
respectively.

The proof is based on a comparison of the string theory expressions
\stringexpr\
for the gauge and gravitational coupling constants with the following field
theoretical
expressions \KL\dWKLL:
\eqn\QFT{
\eqalign{
{1 \over g^2_{\rm grav} (p^2)} & =  {\rm Re} \left( \log F_1^{\rm heterotic}
\right) + {b_{\rm grav} \over 16 \pi^2} \left( \log {M_{\rm Planck}^2 \over
p^2} + K \right) \cr
{1 \over g^2_{\rm gauge} (p^2)} & = {\rm Re} \left( -i \tilde{S} - {1 \over 2
(s
+ 4) \pi^2} \log \Psi_{\rm gauge} \right) + {b_{\rm gauge} \over 16 \pi^2}
\left( \log {M_{\rm Planck}^2 \over p^2} + K \right) , \cr
}
}
where the K\"ahler potential is
\eqn\Kahlerpot{
K = -\log {\rm Re \;} (-i S) - \log \bigl( - (y_2, y_2) \bigr) + {\rm const}
}
and the Planck and string scales are related as $M_{\rm Planck}^2 = M_{\rm
string}^2 {\rm Re \;} (-i S)$. The functions $\log F_1^{\rm heterotic}$ and
$\Psi_{\rm gauge}$ are holomorphic and transform under duality in such a way
that $g_{\rm grav}^{-2} (p^2)$ and $g_{\rm gauge}^{-2} (p^2)$ are invariant.

We start with the equations for $g_{\rm gauge}^{-2} (p^2)$. From the above, it
follows that
\eqn\gaugeinv{
{1 \over 2 (s + 4) \pi^2} {\rm Re} \; \log \Psi_{\rm gauge} + {b_{\rm gauge}
\over
16 \pi^2} \log \bigl( -(y_2, y_2) \bigr)
}
should be invariant under the duality group. We take this quantity to equal the
manifestly invariant ${1 \over 16 \pi^2} \left( {1 \over s + 4} \nabla^2 - 1
\right) \Delta_{\rm gauge}$, where
\eqn\Laplacian{
\nabla^2 = - 2 (y_2, y_2) \left( \eta^{AB} - {2 \over (y_2, y_2)} y_2^A y_2^B
\right) \partial_A \partial_{\bar{B}}
}
is the Laplacian for the metric following from the K\"ahler potential \Kahler\
restricted to the subspace where $\bar{y} \cdot \bar{Q} = 0$. (The tensor
$\eta^{AB}$ is the inverse of $\eta_{AB}$ defined through $\eta_{AB} y^A
y^{\prime B} = (y, y^\prime)$; note that the indices $A$ and $B$ take $17$
different values on this subspace.) It is now a straightforward computation to
verify the above expression for $\Psi_{\rm gauge}$.
This is in fact the unique solution with the correct
singularity structure \HM.  Note that if the quartic tensors \dhatgr\ and
\dhatQ\
in the rational terms had not
vanished, this procedure to construct a holomorphic function with the right
transformation
properties would not have worked.
Comparing the two expressions for $g_{\rm gauge}^{-2} (p^2)$ and using
\eqn\SDelta{
\eqalign{
{1 \over 16 \pi^2} \Delta^{\rm univ} & = {1 \over - (y_2, y_2)} {\rm Re} \left(
{\cal
F}_0^{(1)} - i y_2^A {\partial \over \partial y^A} {\cal F}_0^{(1)} \right) \cr
-i \tilde{S} & = -i S - {1 \over s + 4} {\partial \over \partial y} \cdot
{\partial
\over \partial y} {\cal F}_0^{(1)} , \cr
}
}
we get the following differential equation for the vector multiplet
prepotential ${\cal F}_0^{(1)}$:
\eqn\PDEs{
\eqalign{
{\rm Re} & \left[ -{1 \over s + 4} {\partial \over \partial y} \cdot
{\partial \over \partial y} {\cal F}_0^{(1)} + {1 \over (y_2, y_2)} \left(
{\cal F}_0^{(1)} - i y_2^A
{\partial \over \partial y^A} {\cal F}_0^{(1)} \right) \right] \cr
& = {1 \over 16 \pi^2} \Delta_{\rm gauge} + {1 \over 2 (s + 4) \pi^2} {\rm Re}
\;
\log \Psi_{\rm gauge} + {b_{\rm gauge} \over
16 \pi^2} \log \bigl( -(y_2, y_2) \bigr) + {\rm const} . \cr
}
}
By direct substitution, one checks that a solution is
\eqn\prepotential{
\eqalign{
{\cal F}_0^{(1)} = & - {i \over 32 \pi} (d_{\rm gauge})_{ABC} y^A y^B y^C
+ {\zeta (3) \chi \over 4 \pi^4} \cr
& + {1 \over 16 \pi^4} \sum_{R > 0}{}^{\! {}^\prime} d (R)
{\rm li}_3 \bigl( (R , y) \bigr) .
}
}
Note that this form of
the prepotential as a sum of a cubic tensor, a constant and an infinite sum of
polylogarithms is what one expects when the heterotic theory has a type II
dual.
On the type II side, the coefficients of the cubic tensor, the constant $\chi$
and the
coefficients of the polylogarithms are then interpreted
as the intersection form, the Euler characteristic and the number of rational
curves
of the Calabi-Yau space respectively.

In fact, it is not difficult to see that \PDEs\ would not be satisfied by any
holomorphic
${\cal F}_0^{(1)}$ and $\Psi_{\rm gauge}$ if the quartic tensor \dhatQ\ had not
vanished. (After multiplication by $(y_2, y_2)$, all other terms in \PDEs\ can
be
written as a sum of holomorphic functions in the moduli $y^A$ times at most
quadratic
polynomials in their complex conjugates $y^{*A}$ plus the complex conjugate
terms.
The term in $\Delta_{\rm gauge}$
involving the quartic tensor is not of this form and must therefore vanish.)
Furthermore, the prepotential should be independent of the factor in the gauge
group
that we are considering up to physically irrelevant terms of the form
$(y_2, y_2) \tilde{d}_A y^A$ for some real $\tilde{d}_A$.
(Indeed, on can show that all solutions to \PDEs\ are
physically equivalent in this sense \HM.) Together with analogous arguments for
the
gravitational corrections, these observations show that the conditions
\condone\ and \condtwo\ indeed follow from $N = 2$ supersymmetry.

We now turn to the equations for $g_{\rm grav}^{-2} (p^2)$. Comparing the two
expressions, we get $\log F_1^{\rm heterotic} = 24 (-i S) + \log {\cal
F}_1^{(1)}$ with
\eqn\Fone{
\log {\cal F}_1^{(1)} = {1 \over 2 \pi^2} \sum_{R > 0}{}^{\! {}^\prime}
\tilde{d} (R)
 {\rm li}_1 \bigl( (R, y) \bigr)
}
modulo terms linear in $y$. Here we have used the condition \condtwo.
Finally, we should check that $F_1^{\rm heterotic}$ transforms in the correct
way. Writing $\log F_1^{\rm heterotic} = 24 \left( -i \tilde{S}
- {1 \over 16 \pi^2} \log \Psi_{\rm grav} \right)$, this is equivalent to
\eqn\gravinv{
{1 \over 16 \pi^2} {\rm Re} \; \log \Psi_{\rm grav} + {b_{\rm grav}
\over 16 \pi^2} \log \bigl( - (y_2, y_2) \bigr)
}
being invariant. It is straightforward to check that this equals the manifestly
invariant quantity ${1 \over 16 \pi^2} \left( {1 \over s + 4} \nabla^2 - 1
\right) \Delta_{\rm grav}$, and one verifies the expression for $\Psi_{\rm
grav}$.

\bigskip
\centerline{\bf Acknowledgements}
We would like to thank J. Harvey for participation at
the beginning of this project and for discussions. We would
also like to thank I. Antoniadis and T. Taylor for
a useful discussion and J. Louis, for some useful
corresondence.  M. H. thanks Argonne National
Laboratory for hospitality. G.M. would like to thank
the organizers of Strings '96 for the opportunity
to present these results. He also thanks
the CERN theory group,  the ITP at Santa Barbara,
and the Aspen Center for Physics for hospitality
during the preparation of the manuscript. This research
is supported by DOE grant DE-FG02-92ER40704,
and by a Presidential Young Investigator Award.

\appendix{A}{Poisson resummation and the orbit method}
In this appendix, we will prove that the threshold corrections can be written
as in \Deltadecomp\ with \grterms\ and \Qterms, following the methods
introduced in appendix B of \DKL\ and further developed in \HM.

The first step is to perform a Poisson resummation  of the sums over $m^+$ and
$m^0$ \HM:
\eqn\Poisson{
\sum_{\bar{r}} \sum_{m^+, n^-, m^0, n^0} q^{{1 \over 2} p_L^2} \bar{q}^{{1
\over 2} p_R^2} =  \sum_{\bar{r}} \sum_A {-(y_2, y_2) \over 2 \tau_2 y_2^-}
q^{{1 \over 2} \bar{R} \cdot \bar{R}} \exp {\cal G} .
}
Here
\eqn\Adef{
A = \pmatrix{n^- & j \cr n^0 & p \cr}
}
is summed over all integer $2 \times 2$ matrices and
\eqn\Gdef{
\eqalign{
{\cal G} =  & {\pi (y_2, y_2) \over 2 (y_2^-)^2 \tau_2} | {\cal A} |^2 - 2 \pi
i y^+
\det A + {\pi \over y_2^-} \left( \bar{R} \cdot \bar{y} \tilde{\cal A}  -
\bar{R} \cdot \bar{y}^* {\cal A} \right) \cr
& - {\pi \over 2 y_2^-} n^0 \left( \bar{y} \cdot \bar{y} \tilde{\cal A} -
\bar{y}^* \cdot \bar{y}^* {\cal A} \right) + {i \pi \bar{y}_2 \cdot \bar{y}_2
\over
(y_2^-)^2} \left( n^- + n^0 y^{-*} \right) {\cal A}  ,
}
}
with ${\cal A}  = \pmatrix{1 & y^-} A \pmatrix{\tau \cr 1}$ and $\tilde{\cal A}
= \pmatrix{1 & y^{-*}} A \pmatrix{\tau \cr 1} $.  We can now write
\eqn\Deltares{
\eqalign{
\Delta_{\rm grav} = & \int_{\cal F} {d^2 \tau \over \tau^2_2} \Biggl(
\sum_{\bar{r}} \sum_{a, b} \sum_A \sum_h {1 \over n} e^{2 \pi i {b \over n}
{\bar R}
\cdot \bar{\gamma}} \left( \tilde{c}_{a, b} (h) - {3 c_{a, b} (h) \over \pi
\tau_2} \right) \cr
& \times 2 {- (y_2 , y_2) \over 2 y_2^-} q^{h + {1 \over 2} \bar{R}
\cdot \bar{R}} \exp {\cal G} - \tau_2 b_{\rm grav} \Biggr) \cr
\Delta_{\rm gauge} = & \int_{\cal F} {d^2 \tau \over \tau^2_2} \Biggl(
\sum_{\bar{r}} \sum_{a, b} \sum_A \sum_h {1 \over n} e^{2 \pi i {b \over n}
{\bar R}
\cdot \bar{\gamma}} c_{a, b} (h) \left( {(\bar{R} \cdot \bar{Q})^2 \over
\bar{Q} \cdot \bar{Q}} -  {1 \over 4 \pi \tau_2} \right) \cr
& \times {- (y_2 , y_2) \over 2
y_2^-} q^{h + {1 \over 2} \bar{R} \cdot \bar{R}} \exp {\cal G} - \tau_2 b_{\rm
gauge} \Biggr)  \cr
}
}
with the Fourier coefficients $c_{a, b} (h)$ and $\tilde{c}_{a, b} (h)$ defined
in \Fourier.

A key property is the behavior of the sum over $A$ under modular
transformations: The contributions from the matrix $A^\prime = A V^{-1}$ for $V
= \pmatrix{a & b \cr c & d \cr} \in SL(2, {\bf Z})$ is given by the
contribution from the matrix $A$ if we change $\tau$ to $\tau^\prime = {a \tau
+ b \over c \tau + d}$. We can therefore apply the method of orbits \DKL:
Rather
than integrating the contributions of all integer $2 \times 2$ matrices over
the fundamental region, we pick a representative matrix $A$ from each orbit of
$SL(2, {\bf Z})$ and integrate over the image of the fundamental region under
all $SL(2, {\bf Z})$ transformations $V$ that yield distinct matrices $A
V^{-1}$. In this way we get
\eqn\decomposition{
\eqalign{
\Delta_{\rm grav} & = \Delta_{\rm grav}^{\rm zero} + \Delta_{\rm grav}^{\rm
reg} + \Delta_{\rm grav}^{\rm deg} \cr
\Delta_{\rm gauge} & = \Delta_{\rm gauge}^{\rm zero} + \Delta_{\rm gauge}^{\rm
reg} + \Delta_{\rm gauge}^{\rm deg} , \cr
}
}
where the three terms correspond to contributions from the zero matrix,
matrices with non-zero determinant and non-zero matrices with zero determinant
respectively, and the $b$-terms are included in the last term.

To get the contribution from the zero orbit, we take $A = 0$ and integrate over
the fundamental region ${\cal F} = \{ \tau \; | \; -1/2 < \tau_1 < 1 / 2, \;
\tau_2 > 0,  \; |\tau| > 1 \}$. We thus get $\Delta_{\rm grav}^{\rm zero} =
\int_{\cal F} {d^2 \tau \over \tau_2^2} J_{\rm grav}$ and $\Delta_{\rm
gauge}^{\rm zero} = \int_{\cal F} {d^2 \tau \over \tau_2^2} J_{\rm gauge}$,
where
\eqn\Jdef{
\eqalign{
J_{\rm grav} & = \sum_{\bar{r}} \sum_{a, b} \sum_h {1 \over n} e^{2 \pi i {b
\over n}
{\bar R} \cdot \bar{\gamma}} \left( \tilde{c}_{a, b} (h) - {3 c_{a, b} (h)
\over \pi \tau_2} \right) 2 {- (y_2 , y_2) \over 2 y_2^-} q^{h + {1 \over 2}
\bar{R} \cdot \bar{R}} \cr
J_{\rm gauge} & = \sum_{\bar{r}} \sum_{a, b} \sum_h {1 \over n} e^{2 \pi i {b
\over n}
{\bar R} \cdot \bar{\gamma}}  c_{a, b} (h) \left( {(\bar{R} \cdot \bar{Q})^2
\over \bar{Q} \cdot \bar{Q}} -  {1 \over 4 \pi \tau_2} \right) {- (y_2 , y_2)
\over 2 y_2^-} q^{h + {1 \over 2} \bar{R} \cdot \bar{R}} . \cr
}
}
It follows from modular invariance that $J_{\rm grav}$ and $J_{\rm gauge}$ can
be written as linear combinations of $(E_2 (\tau) - {3 \over \pi \tau_2}) E_4
(\tau) E_6 (\tau) \eta^{-24} (\tau)$, $J (\tau)$ and $1$ with coefficients that
can be determined by considering the coefficients of $q^{-1}$, $q^0
\tau_2^{-1}$ and $q^0$:
\eqn\Jcoeff{
\eqalign{
J_{\rm grav} = & \sum_{\bar{r}} \sum_{a, b} {1 \over n} e^{2 \pi i {b \over n}
{\bar R}
\cdot \bar{\gamma}} c_{a, b} \left(- {1 \over 2} \bar{R} \cdot \bar{R} \right)
{- (y_2 , y_2) \over 2 y_2^-} \cr
& \times \left( - {1 \over 120} \right) (E_2 (\tau) - {3 \over \pi \tau_2}) E_4
(\tau) E_6 (\tau) \eta^{-24} (\tau) \cr
J_{\rm gauge} = & \sum_{\bar{r}} \sum_{a, b} {1 \over n} e^{2 \pi i {b \over n}
{\bar R}
\cdot \bar{\gamma}} c_{a, b} \left(- {1 \over 2} \bar{R} \cdot \bar{R} \right)
{- (y_2 , y_2) \over 2 y_2^-} \biggl( - {7 \over 20} +
{(\bar{R} \cdot \bar{Q})^2 \over \bar{Q} \cdot \bar{Q}} \cr
& \;\;\;\;\;\;\;\;\;\; - {1 \over 2880} (E_2 (\tau) - {3 \over \pi \tau_2}) E_4
(\tau)
E_6 (\tau) \eta^{-24} (\tau) + {1 \over 2880} J (\tau) \biggr) \cr
}
}
The formula
\eqn\modularintegral{
\int_{\cal F} {d^2 \tau \over \tau_2^2} \left(E_2 - {3 \over \pi
\tau_2}\right)^k W = \left. {\pi \over 3 (k + 1)} E_2^{(k + 1)} W
\right|_{q^0},
}
which is valid for an arbitrary meromorphic modular form $W$ of weight $-2 k$
\ref\LSW{
W. Lerche, A. N. Schellekens and N. P. Warner, `Lattices and strings',
{\it Phys. Rep.} {\bf 177} (1989) 1.
},
now gives
\eqn\zerovalues{
\eqalign{
\Delta_{\rm grav}^{\rm zero} & = \sum_{\bar{r}} \sum_{a, b} {1 \over n} e^{2
\pi i
{b \over n} {\bar R} \cdot \bar{\gamma}} c_{a, b} \left(- {1 \over 2} \bar{R}
\cdot
\bar{R} \right) {- (y_2 , y_2) \over 2 y_2^-} {2 \pi \over 5} \cr
\Delta_{\rm gauge}^{\rm zero} & = \sum_{\bar{r}} \sum_{a, b} {1 \over n} e^{2
\pi i
{b \over n} {\bar R} \cdot \bar{\gamma}} c_{a, b} \left(- {1 \over 2} \bar{R}
\cdot
\bar{R} \right) {- (y_2 , y_2) \over 2 y_2^-} {\pi \over 3} \left( {(\bar{R}
\cdot \bar{Q})^2 \over \bar{Q} \cdot \bar{Q}} - {1 \over 20} \right) . \cr
}
}

For the regular orbits, we take $A = \pmatrix{ k & j \cr 0 & p \cr }$ with $0
\leq j < k$ and $p \neq 0$ so that
\eqn\Greg{
\eqalign{
{\cal G} = & {\pi (y_2, y_2) \over 2 (y_2^-)^2 \tau_2} \left|k \tau + j + p y^-
\right|^2 - 2 \pi i y^+ k p \cr
& + {\pi \over y_2^-}  \left( \bar{R} \cdot \bar{y} (k \tau + j + p y^{-*}) -
\bar{R} \cdot \bar{y}^* (k \tau + j + p y^-) \right) \cr
& +  {i \pi \over (y_2^-)^2} \bar{y}_2 \cdot \bar{y}_2 k (k \tau + j + p y^-)
\cr
}
}
and integrate over the double cover of the upper half-plane ${\cal H} = \{ \tau
\; | \; \tau_2 > 0 \}$. We thus get
\eqn\regular{
\eqalign{
\Delta_{\rm grav}^{\rm reg} & = \sum_{\bar{r}} \sum_{a, b} {1 \over n} e^{2 \pi
i
{b \over n} {\bar R} \cdot \bar{\gamma}} 2 \left( \tilde{S}_{1/2} - {3 \over
\pi}
S_{3/2} \right) \cr
\Delta_{\rm gauge}^{\rm reg} & = \sum_{\bar{r}} \sum_{a, b} {1 \over n} e^{2
\pi i
{b \over n} {\bar R} \cdot \bar{\gamma}} \left( {(\bar{R} \cdot \bar{Q})^2
\over
\bar{Q} \cdot \bar{Q}} S_{1/2} - {1 \over 4 \pi} S_{3/2} \right) , \cr
}
}
where
\eqn\Sdef{
\eqalign{
\tilde{S}_{1/2} & = 2 \int_{\cal H} {d^2 \tau \over \tau_2^2} \sum_{k > 0}
\sum_{0 \leq j < k} \sum_{p \neq 0} \sum_h \tilde{c}_{a, b} (h) {- (y_2 , y_2)
\over 2 y_2^-} q^{h + {1 \over 2} \bar{R} \cdot \bar{R}} \exp {\cal G} \cr
S_{1/2} & = 2 \int_{\cal H} {d^2 \tau \over \tau_2^2} \sum_{k > 0} \sum_{0 \leq
j < k} \sum_{p \neq 0} \sum_h c_{a, b} (h) {- (y_2 , y_2) \over 2 y_2^-} q^{h +
{1 \over 2} \bar{R} \cdot \bar{R}} \exp {\cal G} \cr
S_{3/2} & = 2 \int_{\cal H} {d^2 \tau \over \tau_2^3} \sum_{k > 0} \sum_{0 \leq
j < k} \sum_{p \neq 0} \sum_h c_{a, b} (h) {- (y_2 , y_2) \over 2 y_2^-} q^{h +
{1 \over 2} \bar{R} \cdot \bar{R}} \exp {\cal G}. \cr
}
}
After performing the Gaussian integral over $\tau_1$, the only $j$-dependence
of these quantities is through the factor $\exp \left(- 2 \pi i (h + {1 \over
2}
\bar{R} \cdot \bar{R}) {j \over k} \right)$. It follows from the invariance
under $\tau
\rightarrow \tau + 1$ that we only need to consider the case where $h + {1
\over 2} \bar{R} \cdot \bar{R}$ is integral, and the sum over $j$ then amounts
to replacing $h$ by $k l - {1 \over 2} \bar{R} \cdot \bar{R}$ and summing over
all integers $l$. The $\tau_2$ integrals give rise to the Bessel functions
$K_{1/2}$ and $K_{3/2}$, which may be expressed in terms of elementary
functions. Finally, the sums over $p$ may be expressed in terms of the
functions
${\rm li}_1$ and ${\cal P}$ defined in \polylog. In this way we get
\eqn\Svalues{
\eqalign{
\tilde{S}_{1/2} = & {\rm Re} \sum_{k > 0} \sum_{l \in {\bf Z}} 4 \tilde{c}_{a,
b} \left(k l - {1 \over 2} \bar{R} \cdot \bar{R} \right)\cr
& \times {\rm li}_1 \left( \bar{R} \cdot \bar{y}_1 + l y^-_1 + k y^+_1 + i
\left| \bar{R} \cdot \bar{y}_2 + l y^-_2 + k y^+_2 \right| \right) \cr
S_{1/2} = & {\rm Re} \sum_{k > 0} \sum_{l \in {\bf Z}} 4 c_{a, b} \left(k l -
{1 \over 2} \bar{R} \cdot \bar{R} \right) \cr
& \times {\rm li}_1 \left( \bar{R} \cdot \bar{y}_1 + l y^-_1 + k y^+_1 + i
\left| \bar{R} \cdot \bar{y}_2 + l y^-_2 + k y^+_2 \right| \right) \cr
S_{3/2} = & {\rm Re} \sum_{k > 0} \sum_{l \in {\bf Z}} {8 \over - (y_2, y_2)}
c_{a, b} \left(k l - {1 \over 2} \bar{R} \cdot \bar{R} \right)  \cr
& \times {\cal P} \left(\bar{R} \cdot \bar{y}_1 + l y^-_1 + k y^+_1 + i \left|
\bar{R} \cdot \bar{y}_2 + l y^-_2 + k y^+_2 \right| \right) . \cr
}
}

For the degenerate orbits, we take $A = \pmatrix{ 0 & j \cr 0 & p \cr }$ with
$(j, p) \neq (0, 0)$ so that
\eqn\Gdeg{
{\cal G} = {\pi (y_2, y_2) \over 2 (y_2^-)^2 \tau_2} |j + p y^-|^2 + {\pi \over
y_2^-} \left( \bar{R} \cdot \bar{y} (j + p y^{-*}) - \bar{R} \cdot \bar{y}^* (j
+ p y^-) \right)
}
and integrate over the strip ${\cal S} = \{ \tau \; | \; -1/2 < \tau_1 < 1/2,
\; \tau_2 > 0 \}$. The integral over $\tau_1$ amounts to putting $h = - {1
\over 2} \bar{R} \cdot \bar{R}$ and we are left with
\eqn\degenerate{
\eqalign{
\Delta_{\rm grav}^{\rm deg} = & \sum_{\bar{r}} \sum_{a, b}{}^{\! {}^0} {1 \over
n} e^{2 \pi i {b \over n} {\bar R} \cdot \bar{\gamma}} \tilde{c}_{a, b} \left(
- {1 \over
2} \bar{R} \cdot \bar{R} \right) 2 U^0 \cr
& + \sum_{\bar{r}} \sum_{a, b}{}^{\! {}^\prime} {1 \over n} e^{2 \pi i {b \over
n} {\bar
R} \cdot \bar{\gamma}} \tilde{c}_{a, b} \left( - {1 \over 2} \bar{R} \cdot
\bar{R} \right) 2 U^\prime \cr
& + \sum_{\bar{r}} \sum_{a, b} {1 \over n} e^{2 \pi i {b \over n} {\bar R}
\cdot
\bar{\gamma}} c_{a, b} \left( - {1 \over 2} \bar{R} \cdot \bar{R} \right)
\left( - {6 \over \pi} \right) V  \cr
\Delta_{\rm gauge}^{\rm deg}  = & \sum_{\bar{r}} \sum_{a, b}{}^{\! {}^0} {1
\over n} e^{2 \pi i {b \over n} {\bar R} \cdot \bar{\gamma}} c_{a, b} \left( -
{1 \over
2} \bar{R} \cdot \bar{R} \right) {(\bar{R} \cdot \bar{Q})^2 \over \bar{Q} \cdot
\bar{Q}} U^0 \cr
& + \sum_{\bar{r}} \sum_{a, b}{}^{\! {}^\prime} {1 \over n} e^{2 \pi i {b \over
n} {\bar
R} \cdot \bar{\gamma}} c_{a, b} \left( - {1 \over 2} \bar{R} \cdot \bar{R}
\right) {(\bar{R} \cdot \bar{Q})^2 \over \bar{Q} \cdot \bar{Q}} U^\prime \cr
&+ \sum_{\bar{r}} \sum_{a, b} {1 \over n} e^{2 \pi i {b \over n} {\bar R} \cdot
\bar{\gamma}} c_{a, b} \left( - {1 \over 2} \bar{R} \cdot \bar{R} \right)
\left( - {1 \over 4 \pi}  \right) V , \cr
}
}
where
\eqn\UVdef{
\eqalign{
U^0 & = \int_0^\infty {d \tau_2 \over  \tau_2^2} \sum_{(j, p) \neq (0, 0)} {-
(y_2, y_2) \over 2 y^-_2} \exp \left({\pi (y_2, y_2) \over 2 (y_2^-)^2 \tau_2}
| j + p y^-|^2 \right) - \int_{\cal F} {d^2 \tau \over \tau_2} \cr
U^\prime & = \int_0^\infty {d \tau_2 \over  \tau_2^2} \sum_{(j, p) \neq (0, 0)}
{- (y_2, y_2) \over 2 y^-_2} \exp {\cal G} \cr
V & = \int_0^\infty {d \tau_2 \over  \tau_2^3} \sum_{(j, p) \neq (0, 0)} {-
(y_2, y_2) \over 2 y^-_2} \exp {\cal G} . \cr
}
}
The sums in the first two terms in $\Delta_{\rm grav}$ and $\Delta_{\rm gauge}$
are over $a$ such that $\bar{R} \cdot \bar{y} = 0$ and $\bar{R} \cdot \bar{y}
\neq 0$ for generic values of $\bar{y}$ respectively, whereas the sum in the
third term is unrestricted. The contributions from $b_{\rm grav}$ and $b_{\rm
gauge}$ have been included in the first term. The calculation of $U^0$ has been
performed in \DKL. To calculate $U^\prime$ and $V$, one first integrates over
$\tau_2$. The sum over $j$ for $p  = 0$ then directly gives rise to ${\rm
li}_2$ and ${\rm li}_4$ functions, whereas for $p \neq 0$ we can perform the
$j$-sum by means of a Sommerfeld-Watson
transformation   and get ${\rm
li}_1$ and
${\cal P}$ functions. In this manner one finds
\eqn\UVvalues{
\eqalign{
U^0 =  & {\pi \over 3} - \log\bigl(-(y_2, y_2) \bigr) - {\cal K} + 4
\sum_{l > 0} {\rm Re} \; {\rm li_1} (y^- l) \cr
U^\prime = & {2 y_2^- \over \pi} {\rm Re} \; {\rm li}_2 \left({\bar{R}
\cdot \bar{y}_2 \over y^-_2} \right) + 4 \sum_{l \geq 0} {\rm Re} \; {\rm li}_1
\left(\bar{R} \cdot \bar{y} + l y^- - \left[ {\bar{R} \cdot \bar{y}_2 \over
y_2^-} \right] y^-\right) \cr
V = & {- 4 (y_2^-)^3 \over \pi^2 (y_2, y_2)} {\rm Re} \; {\rm li}_4
\left({\bar{R} \cdot \bar{y}_2 \over y_2^-} \right) \cr
& - {8 \over (y_2, y_2)}
\sum_{l \geq 0} {\rm Re} \; {\cal P} \left( \bar{R} \cdot \bar{y} + l y^- -
\left[ {\bar{R}
\cdot \bar{y}_2 \over y_2^-} \right] y^- \right) , \cr
}
}
where  $[ x ]$ denotes the largest integer less than or equal to $x$.

In order to compare with \Deltadecomp, we separate the ${\cal P}$ terms with $l
= 0$ in $\Delta^{\rm deg}_{\rm grav}$ and $\Delta^{\rm deg}_{\rm gauge}$
according to whether $\bar{R} \cdot \bar{y} = 0$ or $\bar{R} \cdot \bar{y} \neq
0$ for generic value of the moduli and note that ${\cal P} (0) = {\zeta (3)
\over 2 \pi}$. We also separate the ${\rm li}_1$ and ${\cal P}$ terms according
to whether $\bar{R}$ is negative or not and use that
\eqn\Rsign{
\sum_{l \geq 0} f \left( \bar{R} \cdot \bar{y} + l y^- - \left[ {\bar{R} \cdot
\bar{y}_2 \over y_2^-} \right] y^- \right) = \cases{
\sum_{l \geq 0} f \bigl( (R ; y) \bigr) & for $\bar{R} \geq 0$; \cr
\sum_{l > 0} f \bigl( (R ; y) \bigr) & for $\bar{R} < 0$; \cr
}
}
for an arbitrary function $f$.

\appendix{B}{The standard embedding}
In this appendix, we will consider the case of the standard embedding of the
spin connection in the gauge group in somewhat more detail. This means that the
shift vector $\bar{\gamma}$ is a root vector of the $E_8 \times E_8$ lattice,
i.e. $\bar{\gamma} \cdot \bar{\gamma} = 2$, which we take to belong to the
first
$E_8$ factor. This is consistent with the level matching condition \levelmatch\
for
$n = 2, 3, 4, 6$, and the constants $k_{a, b}$ are given in the table below:

\vbox{
$$
\halign{\hfil # & \hfil # & \hfil #& \hfil #& \hfil #& \hfil #& \hfil #& \hfil
#& \hfil #& \hfil #
& \hfil #& \hfil #& \hfil #& \hfil # \cr
& ${b \over n} =$ & ${0 \over 12}$ & $\;\;\; {1 \over 12}$ & $\;\;\; {2 \over
12}$ &
$\;\;\; {3 \over 12}$ &
$\;\;\; {4 \over 12}$ & $\;\;\; {5 \over 12}$ & $\;\;\; {6 \over 12}$ & $\;\;\;
{7 \over 12}$ &
$\;\;\; {8 \over 12}$ &
$\;\;\; {9 \over 12}$ & $\;\;\; {10 \over 12}$ & $\;\;\; {11 \over 12}$ \cr
${a \over n} = $ \cr
${0 \over 12}$ & & 0 & & 4 & 16 & 36 & & 64 & & 36 & 16 & 4 \cr
${1 \over 12}$ \cr
${2 \over 12}$ & & 4 & & 4 & & 4 & & 4 & & 4 & & 4 \cr
${3 \over 12}$ & & 16 & & & 16 & & & 16 & & & 16 \cr
${4 \over 12}$ & & 36 & & 4 & & 36 & & 4 & & 36 & & 4 \cr
${5 \over 12}$ \cr
${6 \over 12}$ & & 64 & & 4 & 16 & 4 & & 64 & & 4 & 16 & 4 \cr
${7 \over 12}$ \cr
${8 \over 12}$ & & 36 & & 4 & & 36 & & 4 & & 36 & & 4 \cr
${9 \over 12}$ & & 16 & & & 16 & & & 16 & & & 16 \cr
${10 \over 12}$ & & 4 & & 4 & & 4 & & 4 & & 4 & & 4 \cr
${11 \over 12}$ \cr
}
$$
}

The unbroken
gauge group is generated by the root vectors $\bar{b}$ of the $E_8 \times E_8$
lattice which fulfil ${1 \over n} \bar{b} \cdot \bar{\gamma} \in {\bf Z}$.
For $n = 3, 4, 6$, this means that $\bar{b}$
is one of the $240 + 126$ roots which are orthogonal to $\bar{\gamma}$,
and the gauge group
in this case is $U(1) \times E_7 \times E_8 \times U(1)^4$. For $n = 2$ there
is the
additional posibility that $\bar{b} = \pm \bar{\gamma} $, and the gauge group
is
enhanced to $SU(2) \times E_7 \times E_8 \times U(1)^4$. The first $U(1)$
factor
for $n = 3, 4, 6$ or the $SU(2)$ factor for $n = 2$ arise only in the orbifold
limit,
so for a smooth $K3$-surface the gauge group is $E_7 \times E_8 \times U(1)^4$.
This can then be broken down further by turning on the Wilson line moduli
$\bar{y}$.
To calculate the threshold correction to the gauge coupling constant, there
must
be at least an unbroken $SU(2)$ factor generated by some $\bar{Q}$, which
is either a root vector in the first $E_8$ factor such that  $\bar{\gamma}
\cdot \bar{Q} = 0$ or an arbitrary root vector in the second $E_8$ factor.

In fact, these four orbifold models only differ by being at different points in
the moduli
space of the $K3$-surface. Since these moduli belong to $N = 2$
hypermultiplets,
which do not mix with the vector multiplet moduli, the threshold corrections
should
be independent of $n$. One must be careful at this point, since the answers
{\it are} in fact
{\it different} when $\bar{y} \cdot \bar{\gamma} \neq 0$. The reason is that in
this case,
we turn on the Wilson modulus
corresponding to the extra $U(1)$ or $SU(2)$ factor which only exists in the
orbifold limit, thus freezing the hypermultiplet moduli at that particular
point in the
moduli space of the $K3$-surface and we are exploring inequivalent branches of
the total moduli space. It is worthwhile checking that we do get equivalent
answers
for different $n$ when $\bar{y} \cdot \bar{\gamma} = 0$. To this end, we note
that
the threshold corrections are of the form
\eqn\generalform{
\eqalign{
& \sum_{\bar{r}} \sum_{k, l} \sum_{a, b} {1 \over n} e^{2 \pi i {b \over n}
\bar{R} \cdot \bar{\gamma}}
c_{a, b} \left( - {1 \over 2} \bar{R} \cdot \bar{R} + k l \right)
g \left( \bar{R} \cdot \bar{y}, k y^+ , l y^- \right) \cr
& = \sum_{\bar{r}} \sum_{k, l} \sum_{a, b} {1 \over n} e^{2 \pi i {b \over n}
\bar{R} \cdot \bar{\gamma}}
q^{{1 \over 2} \bar{R} \cdot \bar{R}}
e^{-2 \pi i {a b \over n^2} \gamma^2} \eta^{-20} (\tau) Z^{\rm K3}_{a, b} (
\tau) \Bigr|_{q^{k l}}
g \left( \bar{R} \cdot \bar{y}, k y^+, l y^- \right)
}
}
for some function $g$. Furthermore the quantity $\bar{R} \cdot \bar{y}$ is
invariant under
shifts of $\bar{r}$ by $\bar{\gamma}$. Altogether, this means that the
equivalence would follow
from the $n$-independence of the functions
\eqn\Cfunction{
\sum_j \sum_{a, b} {1 \over n} e^{2 \pi i {b \over n} (\bar{R} + j
\bar{\gamma}) \cdot \bar{\gamma}}
q^{{1 \over 2} (\bar{R} + j \bar{\gamma}) \cdot (\bar{R} + j \bar{\gamma})}
e^{-2 \pi i {a b  \over n^2} \gamma^2} \eta^{-20} (\tau) Z_{a, b}^{\rm K3}
(\tau)
}
for any $\bar{r}$. Performing the sum over $j$ by means of the Jacobi triple
product
identity, using the form \kthree\ of $Z_{a, b}^{\rm K3} (\tau)$ and dropping
some
manifestly $n$-independent factors, we get
\eqn\Dfunction{
\eqalign{
\sum_{a, b} {1 \over n} & k_{a, b}
\left( e^{2 \pi i {b \over n}} q^{{a \over n}} \right)^{(\bar{r} \cdot
\bar{\gamma} + 1)}
\left( 1 - q^{{a \over n}} e^{2 \pi i {b \over n}} \right)^{-2} \cr
\times & \prod_{j = 1}^\infty
\left(1 - q^{j + {a \over n}} e^{2 \pi i {b \over n}} \right)^{-2}
\left(1 - q^{j - {a \over n}} e^{-2 \pi i {b \over n}} \right)^{-2} \cr
& \times
\left( 1 + q^{2 j - 1 + \bar{r} \cdot \bar{\gamma} + 2 {a \over n}} e^{4 \pi i
{b \over n}} \right)
\left( 1 + q^{2 j - 1 - \bar{r} \cdot \bar{\gamma} - 2 {a \over n}} e^{-4 \pi i
{b \over n}} \right) .
}
}
By expanding out the first few terms, one may check that, for integer
$\bar{r} \cdot \bar{\gamma}$, we indeed get the same result for $n = 2, 3, 4,
6$, although
this is not manifest in \Dfunction.

For $n = 2$, it is now straightforward to calculate
\eqn\zerovalues{
\eqalign{
\sum_b {1 \over n} e^{2 \pi i {b \over n} \bar{R} \cdot \bar{\gamma}} c_{a, b}
\left( -
{1 \over 2} \bar{R} \cdot \bar{R} \right) & = 128 \cr
\sum_b {1 \over n} e^{2 \pi i {b \over n} \bar{R} \cdot \bar{\gamma}}
\tilde{c}_{a, b}
\left( - {1 \over 2} \bar{R} \cdot \bar{R} \right) & = -64 \cr
}
}
for $\bar{r} =  a = 0$ and
\eqn\sumvalues{
\eqalign{
\sum_b & {1 \over n} e^{2 \pi i {b \over n} \bar{R} \cdot \bar{\gamma}} c_{a,
b}
\left( - {1 \over 2} \bar{R} \cdot \bar{R} \right) =
\sum_b {1 \over n} e^{2 \pi i {b \over n} \bar{R} \cdot \bar{\gamma}}
\tilde{c}_{a, b}
\left( - {1 \over 2} \bar{R} \cdot \bar{R} \right)\cr \cr
& = \cases{
8 & for $a = 0$, $\bar{r} \cdot \bar{r} = 2$, $\bar{r} \cdot \bar{\gamma} = 0$
\cr \cr
-8 & for $a = 0$, $\bar{r} \cdot \bar{r} = 2$, $\bar{r} \cdot \bar{\gamma} =
\pm 1$ \cr \cr
8 & for $a = 0$, $\bar{r} = \pm \bar{\gamma}$ \cr\cr
-256 & for $a = 1$, $\bar{r} = 0$ or $a = 1$, $\bar{r} = - \bar{\gamma}$ \cr\cr
-64 & for $a = 1$, $\bar{r} \cdot \bar{r} = 2$, $\bar{r} \cdot \bar{\gamma} =
-1$ \cr \cr
0 & otherwise .
}
}
}
The constant $\chi$ receives contributions from $\bar{r}$ and $a$ such that
$\bar{R}$
is parallell to $\gamma$, i.e.
\eqn\chivalue{
\chi = {1 \over 4} \left( 128 + 8 + 8 - 256 -256 \right) = - 92 .
}
This agrees with the expected result $2 \times (19 - 65 ) = - 92$ from $19$
vector multiplets and $20 + 45 = 65$ hypermultiplet moduli for the $K3$-surface
and the gauge bundle.

To calculate the rational terms in the threshold corrections,
we also need some information about the $E_8 \times E_8$ root lattice.
 Let $\bar{v}$ be an arbitrary vector in the vector space of the $E_8$ root
lattice.
Using the rotational symmetry to fourth order of this lattice \ref\PSW{
J. Patera, R. T. Sharp and P. Winternitz, `Higher indices of group
representations',
{\it J. Math. Phys.} {\bf 17} (1976) 1972.
}, we can calculate the following sums over root vectors $\bar{b}$ with
the indicated inner products with $\bar{\gamma}$:
\eqn\rootsums{
\eqalign{
\sum_{\bar{b} \cdot \bar{\gamma} = 0} 1 & = 126 \cr
\sum_{\bar{b} \cdot \bar{\gamma} = 1} 1 & = 56 \cr
\sum_{\bar{b} \cdot \bar{\gamma} = 0} (\bar{b} \cdot \bar{v})^2 &
= 36 \bar{v} \cdot \bar{v} - 18 (\bar{v} \cdot \bar{\gamma})^2 \cr
\sum_{\bar{b} \cdot \bar{\gamma} =1} (\bar{b} \cdot \bar{v})^2 &
= 12 \bar{v} \cdot \bar{v} + 8 (\bar{v} \cdot \bar{\gamma})^2 \cr
\sum_{\bar{b} \cdot \bar{\gamma} = 0} (\bar{b} \cdot \bar{v})^4 &
= 24 (\bar{v} \cdot \bar{v})^2 - 24 \bar{v} \cdot \bar{v} (\bar{v} \cdot
\bar{\gamma})^2
+ 6 (\bar{v} \cdot \bar{\gamma})^4 \cr
\sum_{\bar{b} \cdot \bar{\gamma} = 1} (\bar{b} \cdot \bar{v})^4 &
= 6 (\bar{v} \cdot \bar{v})^2 +12 \bar{v} \cdot \bar{v} (\bar{v} \cdot
\bar{\gamma})^2
- 4 (\bar{v} \cdot \bar{\gamma})^4 . \cr
}
}

We may now calculate (Again using $n = 2$ for simplicity, although the result
would
be the same with $n = 3, 4, 5$ by the above argument):
\eqn\sumone{
\eqalign{
\sum_{\bar{r}} \sum_{a, b} {1 \over n} e^{2 \pi i {b \over n} \bar{R} \cdot
\bar{\gamma}}
c_{a, b} \left( - {1 \over 2} \bar{R} \cdot \bar{R} \right) & = -1920 \cr
\sum_{\bar{r}} \sum_{a, b} {1 \over n} e^{2 \pi i {b \over n} \bar{R} \cdot
\bar{\gamma}}
\tilde{c}_{a, b} \left( - {1 \over 2} \bar{R} \cdot \bar{R} \right) & = -2112 .
\cr
}
}
In the following formulas, we have decomposed
$\bar{y}_2 = \bar{y}_2^1 + \bar{y}_2^2$, where the terms refer to the first and
second $E_8$ factor. We also introduce the sets $r^1_0$ and $r^1_1$ of
positive roots of the first $E_8$ factor whose inner product with
$\bar{\gamma}$ equal
$0$ and $1$ and the set $r^2$ of positive roots of the second $E_8$ factor.
The symbol $\check{c}_{a, b}$ stands for either of the functions $c_{a, b}$ or
$\tilde{c}_{a, b}$.
\eqn\sumtwo{
\eqalign{
\sum_{\bar{r}} \sum_{a, b}  {1 \over n} e^{2 \pi i {b \over n} \bar{R} \cdot
\bar{\gamma}}
\check{c}_{a, b} \left( - {1 \over 2} \bar{R} \cdot \bar{R} \right)
| \bar{R} \cdot \bar{y}_2 |
& = 16 \sum_{\bar{r} \in r^1_0} \bar{r} \cdot \bar{y}^1_2 - 160 \sum_{\bar{r}
\in r^1_1}
\bar{r} \cdot \bar{y}^1_2  \cr
& \;\;\;\; + 16 \sum_{\bar{r} \in r^2} \bar{r} \cdot \bar{y}^2_2 \cr
\sum_{\bar{r}} \sum_{a, b} {1 \over n} e^{2 \pi i {b \over n} \bar{R} \cdot
\bar{\gamma}}
\check{c}_{a, b} \left( - {1 \over 2} \bar{R} \cdot \bar{R} \right) (\bar{R}
\cdot \bar{y}_2)^2
& = -672 \bar{y}^1_2 \cdot \bar{y}^1_2  + 480 \bar{y}^2_2 \cdot \bar{y}^2_2 \cr
\sum_{\bar{r}} \sum_{a, b} {1 \over n} e^{2 \pi i {b \over n} \bar{R} \cdot
\bar{\gamma}}
\check{c}_{a, b} \left( - {1 \over 2} \bar{R} \cdot \bar{R} \right)
| \bar{R} \cdot \bar{y}_2 |^3
& = 16 \sum_{\bar{r} \in r^1_0} (\bar{r} \cdot \bar{y}^1_2)^3 - 160
\sum_{\bar{r} \in r^1_1}
(\bar{r} \cdot \bar{y}^1_2)^3  \cr
& \;\;\;\; + 16 \sum_{\bar{r} \in r^2} (\bar{r} \cdot \bar{y}^2_2)^3 \cr
\sum_{\bar{r}} \sum_{a, b} {1 \over n} e^{2 \pi i {b \over n} \bar{R} \cdot
\bar{\gamma}}
\check{c}_{a, b} \left( - {1 \over 2} \bar{R} \cdot \bar{R} \right) (\bar{R}
\cdot \bar{y}_2)^4
& = -288 (\bar{y}^1_2 \cdot \bar{y}^1_2)^2  + 288 (\bar{y}^2_2 \cdot
\bar{y}^2_2)^2 . \cr
}
}
The sums involving the charge $\bar{Q}$ depend on which $E_8$ factor $\bar{Q}$
belongs to. In the case where $\bar{Q}$ belongs to the first $E_8$ factor,
 we must have
$\bar{Q} \cdot \bar{\gamma} = \bar{y}^1 \cdot \bar{Q} = \bar{y}^1 \cdot
\bar{\gamma} = 0$.
We then get
\eqn\sumthree{
\eqalign{
\sum_{\bar{r}} \sum_{a, b} {1 \over n} e^{2 \pi i {b \over n} \bar{R} \cdot
\bar{\gamma}}
\check{c}_{a, b} \left( - {1 \over 2} \bar{R} \cdot \bar{R} \right) {(\bar{R}
\cdot \bar{Q})^2 \over \bar{Q} \cdot \bar{Q}} & = -672 \cr
\sum_{\bar{r}} \sum_{a, b} {1 \over n} e^{2 \pi i {b \over n} \bar{R} \cdot
\bar{\gamma}}
\check{c}_{a, b} \left( - {1 \over 2} \bar{R} \cdot \bar{R} \right) {(\bar{R}
\cdot \bar{Q})^2 \over \bar{Q} \cdot \bar{Q}} | \bar{R} \cdot \bar{y}_2 |
& = 16 \sum_{\bar{r} \in r^1_0} {(\bar{r} \cdot \bar{Q})^2 \over \bar{Q} \cdot
\bar{Q}}
\bar{r} \cdot \bar{y}^1_2 \cr
& \;\;\;\;  - 160 \sum_{\bar{r} \in r^1_1} {(\bar{r} \cdot \bar{Q})^2 \over
\bar{Q} \cdot \bar{Q}}
\bar{r} \cdot \bar{y}^1_2 \cr
\sum_{\bar{r}} \sum_{a, b} {1 \over n} e^{2 \pi i {b \over n} \bar{R} \cdot
\bar{\gamma}}
\check{c}_{a, b} \left( - {1 \over 2} \bar{R} \cdot \bar{R} \right) {(\bar{R}
\cdot \bar{Q})^2 \over \bar{Q} \cdot \bar{Q}} (\bar{R} \cdot \bar{y}_2)^2
& =  -96 \bar{y}^1_2 \cdot \bar{y}^1_2 . \cr
}
}
In the case where $\bar{Q}$ belongs to the second $E_8$ factor, we must have
$\bar{y}^1 \cdot \bar{Q} = \bar{y}^2 \cdot \bar{\gamma} = 0$. The sums are then
\eqn\sumfour{
\eqalign{
\sum_{\bar{r}} \sum_{a, b} {1 \over n} e^{2 \pi i {b \over n} \bar{R} \cdot
\bar{\gamma}}
\check{c}_{a, b} \left( - {1 \over 2} \bar{R} \cdot \bar{R} \right) {(\bar{R}
\cdot \bar{Q})^2 \over \bar{Q} \cdot \bar{Q}} & = 480 \cr
\sum_{\bar{r}} \sum_{a, b} {1 \over n} e^{2 \pi i {b \over n} \bar{R} \cdot
\bar{\gamma}}
\check{c}_{a, b} \left( - {1 \over 2} \bar{R} \cdot \bar{R} \right) {(\bar{R}
\cdot \bar{Q})^2 \over \bar{Q} \cdot \bar{Q}} | \bar{R} \cdot \bar{y}_2 |
& = 16 \sum_{\bar{r} \in r^2} {(\bar{r} \cdot \bar{Q})^2 \over \bar{Q} \cdot
\bar{Q}}
\bar{r} \cdot \bar{y}^2_2 \cr
\sum_{\bar{r}} \sum_{a, b} {1 \over n} e^{2 \pi i {b \over n} \bar{R} \cdot
\bar{\gamma}}
\check{c}_{a, b} \left( - {1 \over 2} \bar{R} \cdot \bar{R} \right) {(\bar{R}
\cdot \bar{Q})^2 \over \bar{Q} \cdot \bar{Q}} (\bar{R} \cdot \bar{y}_2)^2
& = 96 \bar{y}^2_2 \cdot \bar{y}^2_2 . \cr
}
}

With this information, one may calculate the rational terms in the threshold
corrections
explicitly and in particular verify that the conditions \condone\ and \condtwo\
are indeed
obeyed in the case of the standard embedding.

\listrefs

\bye